\documentclass[preprint,showpacs,floatfix]{revtex4}

\usepackage{amssymb,amsmath,latexsym}
\usepackage{graphicx,bm}

\begin{document}
\title{Spin polarization amplification within nonmagnetic semiconductors at room temperature}
\author{Soon-Wook Jung and Hyun-Woo Lee}
\affiliation{Department of Physics, Pohang University of Science and Technology,
Pohang, Kyungbuk 790-784, Korea}
\date{\today}

\begin{abstract}
We demonstrate theoretically that the spin polarization of current
can be electrically amplified within nonmagnetic semiconductors by
exploiting the fact the spin current, compared to the charge
current, is weakly perturbed by electric driving forces. As a
specific example, we consider a T-shaped current branching
geometry made entirely of a nonmagnetic semiconductor, where the
current is injected into one of the branches (input branch) and
splits into the other two branches (output branches). We show that
when the input current has a moderate spin polarization, the spin
polarization in one of the output branches can be higher than the
spin polarization in the input branch and may reach 100\% when the
relative magnitudes of current-driving electric fields in the two
output branches are properly tuned. The proposed amplification
scheme does not use ferromagnets or magnetic fields, and does not
require low temperature operation, providing an efficient way to
generate a highly spin polarized current in nonmagnetic
semiconductors at room temperature.
\end{abstract}

\maketitle

One of major challenges of the spintronics is to electrically
generate a highly spin polarized current in a nonmagnetic
semiconductor at room temperature \cite{Prinz, Wolf, Zutic}. High
spin polarization (SP) of 40-90\% has been reported
\cite{Fiederling, Jonker1, Park, Jonker2, Dorpe} in electrical
spin injection from a magnetic semiconductor to a nonmagnetic
semiconductor but those reports are yet limited to low
temperatures since the performance of a magnetic semiconductor as
a spin aligner degrades considerably at room temperature. On the
other hand, in electrical spin injection from a conventional
metallic ferromagnet to a nonmagnetic semiconductor, the
temperature dependence is expected to be weaker since the Curie
temperatures of many ferromagnets are higher than the room
temperature. In this case, however, the conductivity mismatch
\cite{Schmidt} between a ferromagnet and a nonmagnetic
semiconductor suppresses the SP of current \cite{Filip} below 1\%
as the current passes the heterojunction interface between a
ferromagnet and a nonmagnetic semiconductor. It has been proposed
\cite{Rashba1, Smith, Fert} that the conductivity mismatch problem
can be resolved by introducing a thin tunnel barrier at the
heterojunction interface. Subsequent experiments \cite{Motsnyi,
Jiang} using oxide tunnel barriers indeed found enhanced SP of
15-30\% at room temperature. Schottky tunnel barriers are also
demonstrated to be effective and the SP of 2-30\% have been
reported \cite{Zhu, Hanbicki} at low temperatures. Though these
results are already encouraging demonstrations of the electrical
spin injection into nonmagnetic semiconductors, further
enhancement is desired for practical semiconductor spintronic
applications. In order to achieve a higher SP at room temperature,
effects of the heterojunction interface on the SP needs
clarification. There are indications \cite{Teresa, Butler} that
the SP is affected not only by the height and width of the
tunnelling barrier but also by electronic structure of the
heterojunction interface.

\begin{figure}
\includegraphics[width=0.80\columnwidth]{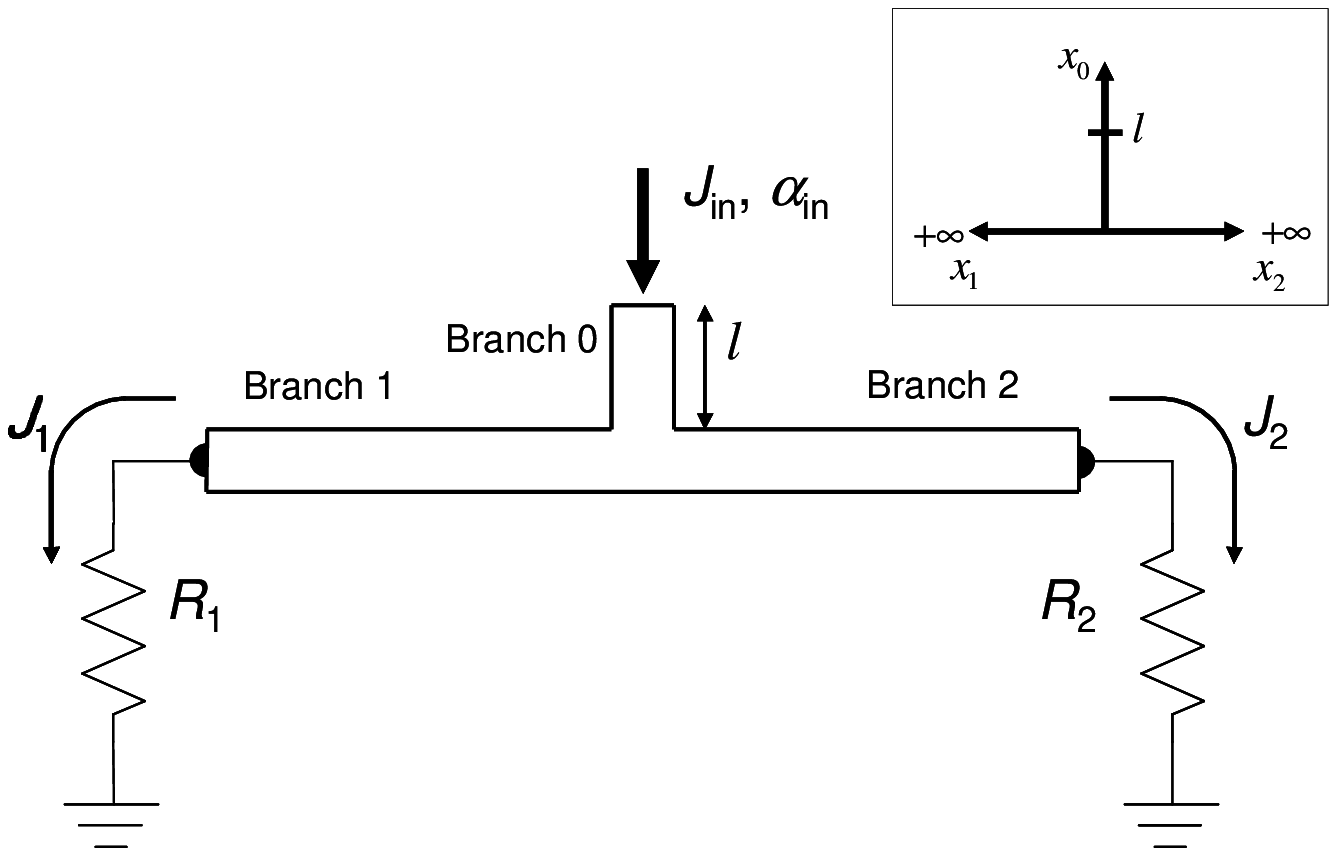}
\caption{Schematic diagram of the branching geometry. The current
$J_{\text{in}}$ with the spin polarization $\alpha_{\text{in}}$ is
injected to the input branch (branch 0) and splitted into the two
output branches (branches 1 and 2) with the branched currents
$J_1$ and $J_2$ ($J_1 + J_2=J_{\text{in}}$). The branching ratios
$\beta_1 \equiv J_1/J_{\text{in}}$, $\beta_2 \equiv
J_2/J_{\text{in}}$ can be modulated by the variable resistances
$R_1$ and $R_2$. When $\beta_1$, $\beta_2$ are properly tuned, the
spin polarization in the branch 1 or 2 can be amplified beyond
$\alpha_{\text{in}}$. Inset: Coordinate system for the branching
geometry. The branching point corresponds to $x_0=x_1=x_2=0$.}
\label{fig1}
\end{figure}

In this paper, we do not address the issue of the interfacial
effects on the SP. Instead we present a method complementary to
the electrical spin injection; namely a method to electrically
amplify, within nonmagnetic semiconductors, the SP of the injected
current. Given spin-up current density $j^\uparrow$ and spin-down
current density $j^\downarrow$, the SP $\alpha$ is defined as
$(j^\uparrow-j^\downarrow)/(j^\uparrow+j^\downarrow)=\Delta j/J$,
where $ \Delta j \equiv j^\uparrow - j^\downarrow $ is the spin
current and $ J \equiv j^\uparrow + j^\downarrow $ is the charge
current. For the SP amplification, we exploit the fact that the
spin current $\Delta j$ is less affected by the electrical driving
forces than the charge current \textit{J} is. One of simple ways
to exploit this is to use a three terminal T-shaped current
branching geometry (Fig.~\ref{fig1}) made entirely of a
nonmagnetic semiconductor. When a charge current $J_{\text{in}}$
with the SP $\alpha_{\text{in}} (<1)$ is injected to the input
branch (branch 0), we demonstrate below that the resulting SP
profile within the current branching geometry does depend on the
relative magnitude of the current-driving electric fields in the
two output branches and that the SP in one of the output branches
can be higher than $\alpha_{\text{in}}$, illustrating the
possibility of the electrical SP amplification.  
The proposed scheme does not need ferromagnets, magnetic
fields, or any heterojunction interfaces, and  it does not require
a low temperature operation. Combined with moderate the SP
achieved via electrical spin injection from a conventional
metallic ferromagnet \cite{Motsnyi, Jiang}, the proposed
amplification scheme provides an efficient tool to generate a
highly spin polarized current in a nonmagnetic semiconductor at
room temperature.
Recently Kato $et al$. have reported the observation of 
electrical spin control in strained semiconductors using the spin-orbit
coupling via strain-induced field\cite{Kato}. 
In contrast, our spin amplification scheme does not require the application of
the strain.

To illustrate our proposal, we use the drift-diffusion model of
electron transport in diffusive semiconductors formulated by Yu
and Flatt\'e \cite{Yu1}. The spin-resolved current densities
$\textbf{j}^\uparrow (\textbf{x},t) $  and $\textbf{j}^\downarrow
(\textbf{x},t)$ can be written as
\begin{subequations}
\label{eq:1}
\begin{align}
\mathbf{j}^\uparrow&=\sigma^\uparrow \mathbf{E} +e D \nabla
n^\uparrow,\label{subeq:1-1}\\
\mathbf{j}^\downarrow&=\sigma^\downarrow \mathbf{E} +e D \nabla
n^\downarrow, \label{subeq:1-2}
\end{align}
\end{subequations}
where $-e(<0)$ is the electron charge, \textit{D} is the diffusion
constant, and $\textbf{E}$ is the electric field. Here
$\sigma^{\uparrow (\downarrow)}$ is the spin-up (spin-down)
conductivity, which depends on the electron density via
$\sigma^{\uparrow (\downarrow)}= \sigma_s/2 + e \nu n^{\uparrow
(\downarrow)}$ where $n^{\uparrow (\downarrow)}$ is the spin-up
(spin-down) electron density deviation from its equilibrium value
$n_s/2$, and $\sigma_s$ is the total conductivity in equilibrium.
Here the mobility $\nu$ is a constant independent of $\textbf{E}$
and $n^{\uparrow (\downarrow)}$, and related to \textit{D} via
$\nu/eD=1/k_B T$ for non-degenerate semiconductors \cite{Yu}.

According to Yu and Flatt\'e \cite{Yu}, a steady state density
profile for a non-degenerate n-doped semiconductor satisfies
\begin{subequations}
\label{eq:2}
\begin{equation}
n^\uparrow + n^\downarrow =0,\label{subeq:2-1}
\end{equation}
\begin{eqnarray}
\nabla^2 (n^\uparrow - n^\downarrow)+ \frac{e \mathbf{E}}{k_B T}
\cdot \nabla(n^\uparrow - n^\downarrow)-\frac{(n^\uparrow -
n^\downarrow)}{L^2_s}=0, \label{subeq:2-2}
\end{eqnarray}
\end{subequations}
where $k_B$ is the Boltzmann constant, \textit{T} is the
temperature, and $L_s$ is the equilibrium spin relaxation length.
The spin resolved electrochemical potentials $\mu^{\uparrow
(\downarrow)}$ in a semiconductor are related\cite{Yu} to the spin
density $n^{\uparrow (\downarrow)}$ via
\begin{equation}
\mu^{\uparrow/ \downarrow}=k_B T \text{ln}\left(1+2 n^{\uparrow
/\downarrow}/n_s\right)+ e \mathbf{E}\cdot \mathbf{x} + C,
\label{chem}
\end{equation}
where \textit{C} is a constant. The charge neutrality condition
Eq.~(\ref{subeq:2-1}) leads to
\begin{subequations}
\label{eq:3}
\begin{equation}
\mathbf{J}=\sigma_s \mathbf{E},\label{subeq:3-1}
\end{equation}
\begin{eqnarray}
\Delta \mathbf{j}=e D \nabla (n^\uparrow - n^\downarrow)+ e \nu
\mathbf{E} (n^\uparrow - n^\downarrow). \label{subeq:3-2}
\end{eqnarray}
\end{subequations}

To get an insight into the electrical SP amplification, it is
useful to first consider briefly the weak electric field limit,
$\lvert \mathbf{E} \rvert \ll E_c \equiv D/\nu L_s=k_B T /e L_s$ ,
where Eqs. (\ref{subeq:2-2}) and (\ref{subeq:3-2}) are simplified
to
\begin{subequations}
\label{sim}
\begin{equation}
\nabla^2(n^\uparrow - n^\downarrow)\approx \frac{(n^\uparrow
-n^\downarrow)}{L^2_s},\label{sim1}
\end{equation}
\begin{eqnarray}
\Delta \mathbf{j}=e D \nabla (n^\uparrow - n^\downarrow).
\label{sim2}
\end{eqnarray}
\end{subequations}
Note that in this limit the spin current $\Delta \textbf{j}$ (and
spin density $ n^\uparrow -n^\downarrow$) is decoupled from the
electric field and governed by diffusion process, while the charge
current \textbf{J} is governed by the electric field
[Eq.~(\ref{subeq:3-1})]. Recalling that the SP $\alpha= \Delta
j/J$, this difference in the coupling strengths of the charge
current and spin current to the electric field opens up the
possibility of the SP amplification via the electric field
modulation within a semiconductor. In a conventional two-terminal
geometry, however, this possibility cannot be exploited 
since the charge conservation fixes the electric field [Eq.~(\ref{subeq:3-1})] 
once the magnitude of the injected current is fixed 
and thus the electric field modulation is not possible.
In the current branching geometry in Fig.~\ref{fig1}, in
contrast, a given injected current $J_{\text{in}}$ is branched
into $J_1$ and $J_2$. Since the charge conservation requires only
the total current conservation $J_1+J_2=J_{\text{in}}$, the
electric field in, say, the branch 1 can be modulated and this
degree of freedom can be exploited to amplify the SP in the branch
1 [See Eq.~(\ref{polarization})].

Next we consider general \textbf{E} and apply the drift-diffusion
model to the T-shaped current branching geometry (Fig.~\ref{fig1})
made of an n-doped diffusive nonmagnetic semiconductor. The charge
current $J_{\text{in}}$ with the SP $\alpha_{\text{in}}$ is
injected into the branch 0 ($0<x_0<l$) and flows to the branch 1
($x_1>0$) and 2 ($x_2>0$) with the branched current $J_1$ and
$J_2$ , respectively. Note that the charge currents $J_i$ and the
electric field $\textbf{E}_i$ $(i=0,1,2)$ are constants within each
branch \textit{i} in the steady state. Taking the coordinates in
the branch \textit{i} to be positive $x_i$ (inset in
Fig.~\ref{fig1}), $J_i$ in our convention is positive (negative)
when the current is flowing outward (inward) from (towards) the
branching point, $x_0=x_1=x_2=0$. Note that $J_0=-J_{\text{in}}$
in this convention. To be specific, we consider below the
\textit{electron injection} into the branch 0. Thus
$J_{\text{in}}<0$ and $J_0=-J_{\text{in}}>0$. And the direction of
the electric field is parallel to the $x_0$-axis in the branch 0
($\textbf{E}_0=E_0 \hat{\textbf{x}}_0$, $E_0>0$), and antiparallel
to the $x_{1(2)}$-axis in the branch 1(2)
($\textbf{E}_{1(2)}=-E_{1(2)} \hat{\textbf{x}}_{1(2)}$,
$E_{1(2)}>0 $). 

The charge current conservation leads to
$J_0=-(J_1+J_2)$. Defining the branching ratios $\beta_1 \equiv
J_1/J_{\text{in}}$ and $\beta_2 \equiv J_2/J_{\text{in}}$, we
impose the branching conditions,
\begin{equation}
\label{cond} \beta_1+\beta_2=1 \quad\text{and}\quad 0\leqq
\beta_1, \beta_2 \leqq 1.
\end{equation}
Due to Eq.~(\ref{eq:2}), the electron density $n^{\uparrow
/\downarrow}_i$ in the branch \textit{i} becomes
\begin{subequations}
\label{solution}
\begin{align}
n^{\uparrow/\downarrow}_0(x_0)&=\pm\bigl(A_0 e^{x_0/L_{d0}}+ B_0
e^{-x_0/L_{u0}}\bigr),\label{sol0}\\
n^{\uparrow/\downarrow}_1(x_1)&=\pm\bigl(A_1 e^{x_1/L_{u1}}+ B_1
e^{-x_1/L_{d1}}\bigr), \label{sol1}\\
n^{\uparrow/\downarrow}_2(x_2)&=\pm\bigl(A_2 e^{x_1/L_{u2}}+ B_2
e^{-x_2/L_{d2}}\bigr), \label{sol2}
\end{align}
\end{subequations}
and the electrochemical potential $\mu_i^{\uparrow/\downarrow}$
becomes
\begin{subequations}
\label{chemsol}
\begin{align}
\mu_0^{\uparrow /\downarrow}&=k_B T \text{ln}\left(1+2
n_0^{\uparrow /\downarrow}/n_s\right)+ e E_0 x_0 +
C_0,\label{ch0}\\
\mu_1^{\uparrow /\downarrow}&=k_B T \text{ln}\left(1+2
n_1^{\uparrow /\downarrow}/n_s\right)- e E_1 x_1 + C_1,
\label{ch1}\\
\mu_2^{\uparrow /\downarrow}&=k_B T \text{ln}\left(1+2
n_2^{\uparrow /\downarrow}/n_s\right)- e E_2 x_2 + C_2,
\label{ch2}
\end{align}
\end{subequations}
where  $L_{ui}=L_s/\Gamma_{ui}$ and $L_{di}=L_s/\Gamma_{di}$ are
the up-stream and down-stream spin diffusion lengths,
respectively\cite{Yu}, with $L_{ui} L_{di}=L_s^2$. Here
$\Gamma_{ui}$ and $\Gamma_{di}$ are given by
\begin{subequations}
\label{length}
\begin{align}
\Gamma_{di}&=-\frac{1}{2}\frac{E_i}{E_c}+ \sqrt{\left(\frac{1}{2}
\frac{E_i}{E_c}\right)^2+1}, \label{ld}\\
\Gamma_{ui}&=\frac{1}{2}\frac{E_i}{E_c}+ \sqrt{\left(\frac{1}{2}
\frac{E_i}{E_c}\right)^2+1}. \label{lu}
\end{align}
\end{subequations}
Note that $L_{di}>L_s>L_{ui}$ since $E_i>0$ is assumed. On the
other hand, when electrons are extracted from the branch 0 and
$E_i<0$, Eqs.~(\ref{solution})-(\ref{length}) remain the same but
$L_{di}<L_s<L_{ui}$. Here we have assumed that the thickness of
each branch is much smaller than the spin relaxation length, so
that the system is essentially one-dimensional. For simplicity, we
also assume that the cross sections of all three branches are the
same. Then $E_0=E_1+E_2$. Determination of the coefficients $A_i$,
$B_i$, and $C_i$ requires the boundary conditions at the branching
point ($x_0=x_1=x_2=0$) and at the end points of the branches
($x_0=l, x_1=\infty, x_2=\infty$). When the three branches make an
ohmic contact with each other (in this case the contact is not
spin-selective), the spin-resolved electrochemical potentials
$\mu^{\uparrow/\downarrow}$ are continuous
\begin{equation}
\mu^{\uparrow /\downarrow}_0(x_0=0)=\mu^{\uparrow
/\downarrow}_1(x_1=0)=\mu^{\uparrow /\downarrow}_2(x_2=0),
\label{bound1}
\end{equation}
and the spin currents are conserved
\begin{equation}
\Delta j_0(x_0=0)+\Delta j_1(x_1=0)+\Delta j_2(x_2=0)=0,
\label{bound2}
\end{equation}
at the branching point. At the end point of the branch 0,
\begin{equation}
\Delta j_0(x_0=l)=\alpha_{\text{in}} J_0, \label{bound3}
\end{equation}
and at the end points of the output branches (\textit{i}=1, 2),
the finiteness of the spin relaxation length imposes,
\begin{equation}
j^\uparrow_i(x_i=\infty)-j^\downarrow_i(x_i=\infty)=0.
\label{bound4}
\end{equation}
Without loss of generality, we may set $C_1=0$. Then the remaining
8 coefficient are fixed by the 8 constraints from
Eqs.~(\ref{bound1})-(\ref{bound4}). After some algebra we obtain
the SP $\alpha_i(x_i)=\Delta j_i(x_i)/J_i$. For example,
$\alpha_1(x_1)$ is given by
\begin{widetext}
\begin{equation}
\alpha_1(x_1)=\frac{\alpha_{\text{in}}}{\beta_1}\frac{
(J_1-J_c\Gamma_{d1})(\Gamma_{u0}+\Gamma_{d0})e^{-x_1/L_{d1}}
}{[(J_{\text{in}}-J_c\Gamma_{d0})(\Gamma_{u0}+\Gamma_{d1}+\Gamma_{d2})e^{\delta\Gamma_{d0}}+
(J_{\text{in}}+J_c\Gamma_{u0})(\Gamma_{d0}-\Gamma_{d1}-\Gamma_{d2})e^{-\delta\Gamma_{u0}}]},
\label{SP}
\end{equation}
\end{widetext}
where $\delta \equiv l/L_s$ and $J_c\equiv \sigma_s E_c$.
This equation is the main result of this paper.

To understand implications of Eq.~(\ref{SP}), we first examine the
small injection current limit, $\vert J_{\text{in}} \vert \ll
J_c$, which is equivalent to the weak electric field limit
addressed briefly above. In this limit, Eq.~(\ref{SP}) is
simplified to
\begin{equation}
\alpha_1(x_1)=\frac{\alpha_{\text{in}}}{\beta_1}
\frac{\text{exp}(-x_1/L_s)}{\text{sinh}(l/L_s)+
2\text{cosh}(l/L_s)}. \label{polarization}
\end{equation}
Note that $\alpha_1(x_1)$ becomes larger than $\alpha_{\text{in}}$
for sufficiently small $\beta_1=J_1/J_{\text{in}}=E_1/E_0$ or for
sufficiently  small $E_1$. Thus a proper tuning of the electric
field $E_1$ in the output branch 1 indeed accomplishes the SP
amplification. For $x_1, l \ll L_s$, the SP amplification occurs
for $\beta_1<1/2$  or $E_1<E_0/2$. In particular when $E_1$ is
tuned so that $\beta_1=\beta^*_1$, where
\begin{equation}
\beta^*_1=\frac{\alpha_{\text{in}}}{\text{sinh}(l/L_s)+
2\text{cosh}(l/L_s)}, \label{value}
\end{equation}
the SP in the branch 1 becomes 100\% at $x_1=0$ and remains close
to 100\% over the length segment of order $L_s$ in the branch 1.
The magnitude of the 100\% spin polarized current is given by
$\beta^*_1 J_{\text{in}}$, which becomes $\alpha_{\text{in}}
J_{\text{in}}/2$ when $l\ll L_s$. Here the factor 2 is due to the
branching of the spin current into the two output branches. The
expression for $j_1^{\uparrow/ \downarrow}$ in the weak field
limit is illustrative;
\begin{equation}
j_1^{\uparrow/ \downarrow}(x_1)=-\frac{1}{2} \sigma_s E_1 \pm
\frac{1}{2} \frac{\alpha_{\text{in}} J_{\text{in}}
\text{exp}(-x_1/L_s)}{\text{sinh}(l/L_s)+ 2
\text{cosh}(l/L_s)}.\label{resolved}
\end{equation}
Note that the spin current $\Delta j_1= j_1^\uparrow -
j_1^\downarrow$ is independent of $E_1$ while the charge current
$J_1= j_1^\uparrow + j_1^\downarrow$ is directly proportional to
$E_1$, thus enabling the SP amplification by the electric field.
We remark that Eqs.~(\ref{polarization}) and (\ref{resolved}) can
be obtained also by using the diffusion equation\cite{Schmidt,
Fert, Rashba2} $\partial^2(\mu^\uparrow - \mu^\downarrow)/\partial
x^2 - (\mu^\uparrow - \mu^\downarrow)/L^2_s=0$, which is used to
describe highly degenerated metal systems.

\begin{figure}
\includegraphics[width=0.80\columnwidth]{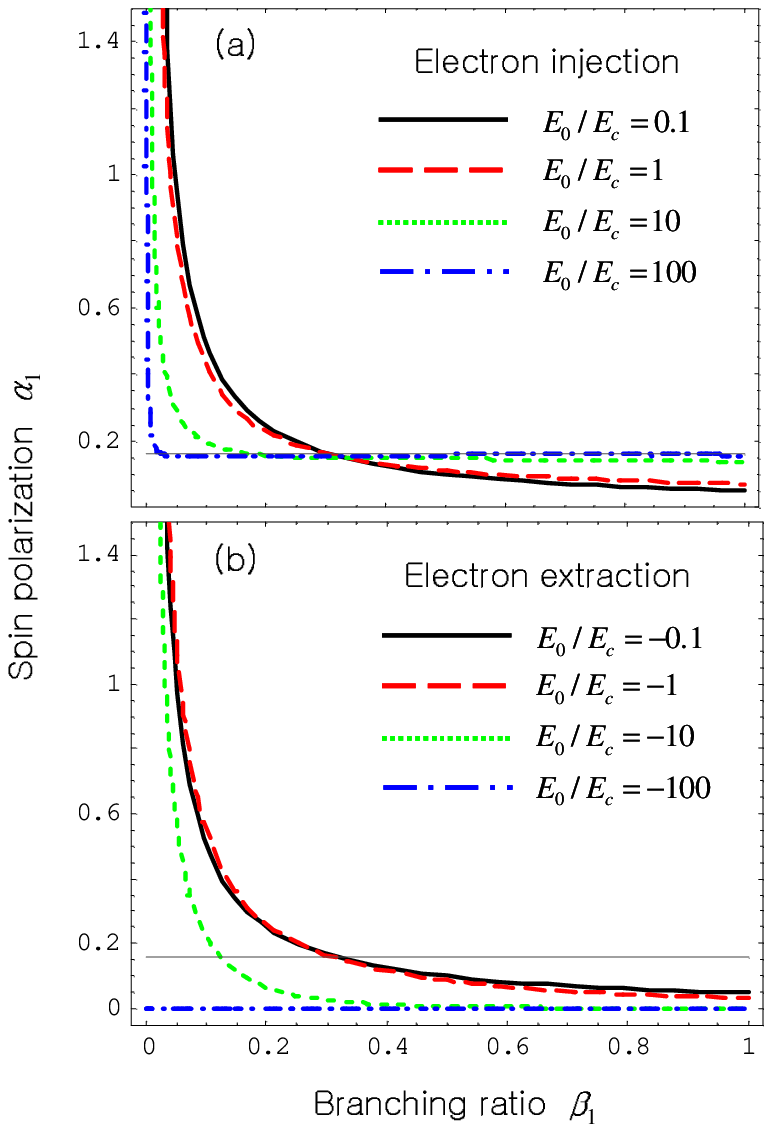}
\caption{(Color online) The spin polarization of the current in
the branch 1 at $x_1=0.3 L_s$ (Fig.~\ref{fig1}) as function of the
branching ratio $\beta_1=J_1/J_{\text{in}}=E_1/E_0$. Here the
injection polarization $\alpha_{\text{in}}=0.16$ [horizontal solid
lines in (a) and (b)] and $l=0.3 L_s$ are assumed. (a) For
electron injection into the branch 0 ($J_{\text{in}}<0$) with
$E_0/E_c=0.1, 1, 10, 100$. (b) For electron extraction from the
branch 0 ($J_{\text{in}}>0$) with $E_0/E_c=-0.1, -1, -10, -100$.}
\label{fig2}
\end{figure}

Figure 2 shows the SP $\alpha_1$ in the branch 1 for various $E_0$
as a function of the branching ratio
$\beta_1=J_1/J_{\text{in}}=E_1/E_0$. For the plots,
$\alpha_{\text{in}}=0.16$, $x_1=0.3 L_s$, and $l=0.3 L_s$ are
used. Note that $\alpha_1(x_1=0.3 L_s)$ is higher than
$\alpha_{\text{in}}$ (horizontal solid lines in Fig.~\ref{fig2})
when $\beta_1$ is smaller than a critical value that depends on
$E_0$, and reaches 1 when $\beta_1$ is reduced further. The SP
amplification for small $\beta_1$ (or small $\lvert E_1 \rvert$)
is most effective for small and moderate injection current
$J_{\text{in}}$ (or for $\lvert E_0 \rvert \lesssim E_c$) and
becomes less effective in the high injection current limit (or for
$\lvert E_0 \rvert \gg E_c$). But even for $\lvert E_0 \rvert \gg
E_c$, the SP amplification is still possible provided that the
branching ratio $\beta_1$ is sufficiently small so that $E_1\ll
E_c$. For $\lvert E_0 \rvert \gg E_c$, $\lvert E_1 \rvert \ll
E_c$, and $l \lesssim L_s$, Eq.~(\ref{SP}) reduces to
$\alpha_1(x_1)\simeq \alpha_{\text{in}}(J_c/\beta_1 \lvert
J_{\text{in}} \rvert) \text{exp}(-x_1/L_s)$ for $J_{\text{in}}<0$
(electron injection into the branch 0) and to $\alpha_1(x_1)\simeq
\alpha_{\text{in}}(J_{\text{in}}/\beta_1 J_c) \text{exp}\lbrace
-[x_1/L_s+ (l/L_s)(J_{\text{in}}/J_c)]\rbrace$ for
$J_{\text{in}}>0$ (electron extraction from the branch 0), which
can be large than $\alpha_{\text{in}}$ for sufficiently small
$\beta_1$. The dependence on the sign of $J_{\text{in}}$ arises
since not only the charge current but also the spin current is
coupled to the electric field in the strong electric field limit.
The coupling between the spin current and the electric field
arises from the drift terms [second term in Eq.~(\ref{subeq:2-2})
and the last term in Eq.~(\ref{subeq:3-2})], and makes the
relaxation of the spin current dependent on the electric field
direction\cite{Yu}.

Next we estimate the field strength $E_1$ for the SP amplification
in real semiconductors at room temperature. For a nondegenerate
n-doped nonmagnetic semiconductor\cite{Fert, Kikkawa2} with the
doping density $n_s=10^{16}\text{cm}^{-3}$, the mobility
$\nu=5400\text{cm}^2/\text{Vs}$, and the equilibrium spin
diffusion length $L_s=1.83\mu \text{m}$, one finds
$E_c=141\text{V}/\text{cm}$ at 300\text{K} and
$J_c=1220\text{A}/\text{cm}^2$. For $E_0=E_c$ and $l=0.3 L_s$, the
SP $\alpha_1$ at $x_1=0.3 L_s$ is lager than $\alpha_{\text{in}}$
for $E_1<0.31 E_c \simeq 44\text{V}/\text{cm}$ and than $5
\alpha_{\text{in}}$ for $E_1<0.046 E_c \simeq
6.5\text{V}/\text{cm}$.

Lastly we comment on several prior proposals to generate highly
spin polarized current by using multiple-terminal structures
\cite{Shao, Long, Wang, Benjamin, Kiselev, Bychkov}. Though the
proposed structures are similar to Fig.~\ref{fig1} in the sense
that they all use structures with multiple terminals, there are
notable differences; The three-terminal structure in
Ref.~\onlinecite{Shao} includes two ferromagnetic electrodes and
three heterojunction interfaces. The three-terminal structures in
Refs.~\onlinecite{Long} and \onlinecite{Wang} use the Coulomb
blockade effect in quantum dots, which limits their operation to
low temperatures. The three-terminal structure in
Ref.~\onlinecite{Benjamin} contains a superconducting electrode,
which again limits its operation to low temperatures. The
ballistic three-terminal structures in Ref.~\onlinecite{Kiselev}
exploit the Rashba spin-orbit coupling \cite{Bychkov} in a
two-dimensional electron gas to generate a highly spin polarized
current. When parameters for InAs/InGaAs heterostructures are
used, it turns out that the operation of this mechanism is limited
to a rather narrow energy range, whose width is an order of
magnitude smaller than the thermal energy at the room temperature.
It is also demonstrated \cite{Yamamoto} that this mechanism
becomes ineffective in the diffusive regime.

In summary, we have demonstrated that the spin polarization can be electrically amplified
within a nonmagnetic semiconductor by using a current branching
geometry. The proposed amplification scheme does not require a
ferromagnet, a magnetic field, or a low temperature operation, and
is thus expected to be an efficient method to generate a highly
spin polarized current in a nonmagnetic semiconductor at room
temperature.

We thank Hu-Jong Lee and Jae-Hoon Park for valuable comments. This
work was supported by the SRC/ERC program (Grant No. R11-2000-071)
and the Basic Research Program (Grant No. R01-2005-000-10352-0) of
MOST/KOSEF and by the Korea Research Foundation Grant (Grant No.
KRF-2005-070-C00055) funded by the Korean Government (MOEHRD).

\begin{thebibliography}{99}
\bibitem{Prinz} G. A. Prinz, Science {\bf 282}, 1660 (1998).
\bibitem{Wolf} S. A. Wolf, D. D. Awschalom, R. A. Buhrman, J. M. Daughton,
S. von Mon$\acute{\text{a}}$r, M. L. Roukes, A. Y. Chtchelkanova,
and D. M. Treger,  Science {\bf 294}, 1488 (2001).
\bibitem{Zutic} I. Zutic, J. Fabian, and S. D. Sarma, Rev. Mod. Phys. {\bf 76}, 323 (2004).
\bibitem{Fiederling} R. Fiederling, M. Keim, G. Reuscher, W. Ossau, G. Schmidt, A. Waag, and L. W. Molenkamp, Nature {\bf 402}, 787 (1999).
\bibitem{Jonker1} B. T. Jonker, Y. D. Park, B. R. Bennett, H. D. Cheong, G. Kioseoglou, and A. Petrou, Phys. Rev. B {\bf 62}, 8180 (2000).
\bibitem{Park} Y. D. Park, B. T. Jonker, B. R. Bennett, G. Itskos, M. Furis, G. Kioseoglou, and A. Petrou
 Appl. Phys. Lett. {\bf 77}, 3989 (2000).
\bibitem{Jonker2} B. T. Jonker, A. T. Hanbicki, Y. D. Park, G. Itskos, M. Furis, G. Kioseoglou, A. Petrou, and X. Wei, Appl. Phys. Lett. {\bf 79}, 3098 (2001).
\bibitem{Dorpe} P. Van Dorpe, Z. Liu, W. Van Roy, V. F. Motsnyi, M. Sawicki, G. Borghs, and J. De Boeck, Appl. Phys. Lett. {\bf 84}, 3495 (2004).
\bibitem{Schmidt} G. Schmidt, D. Ferrand, L. W. Molenkamp, A.
T. Filip, and B. J. van Wees, Phys. Rev. B {\bf 62}, R4790 (2000).
\bibitem{Filip} A. T. Filip, B. H. Hoving, F. J. Jedema, B. J. van Wees, B. Dutta, and S. Borghs, Phys. Rev. B {\bf 62},9996 (2000).
\bibitem{Rashba1} E. I. Rashba, Phys. Rev. B {\bf 62}, R16267 (2000).
\bibitem{Smith} D. L. Smith and R. N. Silver, Phys. Rev. B {\bf 64}, 045323 (2001).
\bibitem{Fert}  A. Fert and H. Jaffres, Phys. Rev. B {\bf 64}, 184420 (2001).
\bibitem{Motsnyi}  V. F. Motsnyi, P. Van Dorpe, W. Van Roy, E. Goovaerts,
V. I. Safarov, G. Borghs, and J. De Boeck, Phys. Rev. B {\bf 68},
245319 (2003).
\bibitem{Jiang} X. Jiang, R. Wang, R. M. Shelby, R. M. Macfarlane, S. R. Bank, J. S. Harris, and S. S. P. Parkin, Phys. Rev. Lett. {\bf 94}, 056601 (2005).
\bibitem{Zhu} H. J. Zhu, M. Ramsteiner, H. Kostial, M. Wassermeier, H.-P. Sch$\ddot{\text{o}}$nherr, and K. H. Ploog, Phys. Rev. Lett. {\bf 87}, 016601 (2001).
\bibitem{Hanbicki} A. T. Hanbicki, O. M. J. van 't Erve, R. Magno, G. Kioseoglou, C. H. Li, B. T. Jonker, G. Itskos, R. Mallory, M. Yasar, and A. Petrou, Appl. Phys. Lett. {\bf 82}, 4092 (2003).
\bibitem{Teresa} J. M. De Teresa, A. Barth$\acute{\text{e}}$l$\acute{\text{e}}$my, A. Fert, J. P. Contour, F. Montaigne, and P.
Seneor, Science {\bf 286}, 507 (1999).
\bibitem{Butler} W. H. Butler, X.-G. Zhang, T. C. Schulthess, and J. M.
MacLaren, Phys. Rev. B {\bf 63}, 054416 (2001).
\bibitem{Kato} Y. Kato, R. C. Myers, A. C. Gossard, and D. D.
Awschalom, Nature(London) {\bf 427} 50 (2004).
\bibitem{Yu1} Z. G. Yu and M. E. Flatt\'e, Phys. Rev. B {\bf 63},
R201202 (2002).
\bibitem{Yu} Z. G. Yu and M. E. Flatt$\acute{\text{e}}$, Phys. Rev. B {\bf 66},
235302 (2002).
\bibitem{Rashba2} E. I. Rashba, Eur. Phys. J. B, {\bf 29}, 513
(2002).
\bibitem{Kikkawa2} J. M. Kikkawa and D. D. Awschalom, Phys. Rev. Lett. {\bf 80}, 4313 (1998).
\bibitem{Shao} L. B. Shao and D. Y.
Xing, Phys. Rev. B {\bf 70}, 201205 (2004).
\bibitem{Long} W. Long, Q.-F. Sun, H.
Guo, and J. Wang, Appl. Phys. Lett. {\bf 83}, 1397 (2003).
\bibitem{Wang} J. Wang,
K. S. Chan, and D. Y. Xing, Phys. Rev. B {\bf 72}, 115311 (2005).
\bibitem{Benjamin} C. Benjamin and R. Citro, Phys. Rev. B {\bf 72}, 085340 (2005).
\bibitem{Kiselev} A. A.
Kiselev and K. W. Kim, Appl. Phys. Lett. {\bf 78}, 775 (2001); J.
Appl. Phys. {\bf 94}, 4001 (2003).
\bibitem{Bychkov} Yu. A. Bychkov and E. I. Rashba, Sov.
Phys. JETP Lett. {\bf 39}, 78 (1984).
\bibitem{Yamamoto} M. Yamamoto, T. Ohtsuki, and
B. Kramer, Phys. Rev. B {\bf 72}, 115321 (2005).
\end{thebibliography}
\end{document}